\documentclass[twocolumn]{aastex631}

\usepackage{graphicx}
\usepackage{amssymb}
\usepackage{amsmath}

\received{May 20, 2024}
\revised{July 17, 2024}
\accepted{August 8, 2024}

\shorttitle{Galaxy assembly bias in the SHMR for red central galaxies from SDSS}
\shortauthors{Oyarz\'un et al.}
\graphicspath{{./}{figures/}}

\begin{document}

\title{Galaxy assembly bias in the stellar-to-halo mass relation for red central galaxies from SDSS}

\email{goyarzu1@jhu.edu}

\author{Grecco A. Oyarz\'un}
\affiliation{The William H. Miller III Department of Physics \& Astronomy, Johns Hopkins University, Baltimore, MD 21218, USA}

\author{Jeremy L. Tinker}
\affiliation{Center for Cosmology and Particle Physics, Department of Physics, New York University, New York, USA, 10003}

\author{Kevin Bundy}
\affiliation{University of California Observatories - Lick Observatory, University of California, Santa Cruz, CA 95064, USA}

\author{Enia Xhakaj}
\affiliation{High-Energy Physics Division, Argonne National Laboratory, 9700 S Cass Ave, Lemont, IL 60439, USA}

\author{J. Stuart B. Wyithe}
\affiliation{Research School of School of Astronomy and Astrophysics, Australian National University, Mt Stromlo, Act 2611, Australia}
\affiliation{ARC Centre of Excellence for All Sky Astrophysics in 3 Dimensions (ASTRO 3D), Australia}



\begin{abstract}
We report evidence of galaxy assembly bias --- the correlation between galaxy properties and biased secondary halo properties at fixed halo mass (M$_\mathrm{H}$) --- in the stellar-to-halo mass relation (SHMR) for red central galaxies from the Sloan Digital Sky Survey. In the M$_\mathrm{H} = 10^{11.5}-10^{13.5}~h^{-1}$~M$_{\odot}$ range, central galaxy stellar mass (M$_*$) is correlated with the number density of galaxies within $10~h^{-1}$~Mpc ($\delta_{10}$), a common proxy for halo formation time. This galaxy assembly bias signal is also present when M$_\mathrm{H}$, M$_*$, and $\delta_{10}$ are substituted with group luminosity, galaxy luminosity, and metrics of the large-scale density field. To associate differences in $\delta_{10}$ with variations in halo formation time, we fitted a model that accounts for (1) errors in the M$_\mathrm{H}$ measured by the \citet{tinker2021,tinker2022} group catalog and (2) the level of correlation between halo formation time and M$_*$ at fixed M$_\mathrm{H}$. Fitting of this model yields that (1) errors in M$_\mathrm{H}$ are $\sim 0.15$~dex and (2) halo formation time and M$_*$ are strongly correlated (Spearman's rank correlation coefficient $\sim 0.85$). At fixed M$_\mathrm{H}$, variations of $\sim 0.4$~dex in M$_*$ are associated with $\sim 1-3$~Gyr variations in halo formation time and in galaxy formation time (from stellar population fitting; \citealt{oyarzun2022}). These results are indicative that halo properties other than M$_\mathrm{H}$ can impact central galaxy assembly.
\end{abstract}



\section{Introduction} \label{1}

In the current cosmological picture, the growth of dark-matter halos and the formation of galaxies are closely connected (e.g. \citealt{white-rees1978,blumenthal1984,mo-white1996,wechsler-tinker2018,vogelsberger2020}). As halo mass (M$_\mathrm{H}$) increases, so does the gravitational pull of the halo, facilitating the accretion of gas, fueling star-formation, and promoting the build-up of stellar mass (M$_*$). The resulting correlation between M$_\mathrm{H}$ and M$_*$ is known as the stellar-to-halo mass relation of central galaxies (SHMR; e.g. \citealt{moster2013}).

While M$_\mathrm{H}$ emerges as the primary halo property driving galaxy growth, simulations of galaxy formation indicate that halo properties other than M$_\mathrm{H}$ also play a role. One of these properties is thought to be halo formation time  (e.g. \citealt{gao2005,tojeiro2017}). Variations in halo assembly history lead to variations in halo gravitational potential with time, ultimately impacting the M$_*$ assembly history of the galaxies that they host (e.g. \citealt{wechsler2002,wechsler2006,gao2005,li2008}). Similarly, halos of different concentrations feature different radial dark-matter density profiles, affecting the assembly histories of halos and galaxies alike, according to simulations (\citealt{ludlow2014,matthee2017,xu2020}). These correlations between galaxy properties and biased secondary halo properties at fixed M$_\mathrm{H}$ that are apparent in theoretical work are known as \textit{galaxy assembly bias} (e.g. \citealt{croton2007,mao2017,zehavi2018,hadzhiyska2021,hadzhiyska2023,montero-dorta2024}).

The major role played by dark-matter halos in simulations of galaxy formation has encouraged observational studies to search for signatures of galaxy-halo co-evolution. For instance, \citet{obuljen2020} found the power spectrum of galaxies in the Baryon Oscillation Spectroscopic Survey of SDSS-III (BOSS; \citealt{dawson2013}) to vary with galaxy central stellar velocity dispersion. These variations --- of the order of $5\sigma$ --- appear at spatial scales that extend beyond the virial radii of dark-matter halos, indicating that galaxy properties correspond with the distribution of halos in the large scale structure. In a separate study, \citet{wang2022} found that the spatial distribution of galaxies in the Sloan Digital Sky Survey (SDSS; \citealt{dr12}) was best reproduced with halo occupation models (e.g. \citealt{tinker2008}) that account for galaxy assembly bias in halo concentration. The same conclusion was reached by \citet{yuan2021} in BOSS and by \citet{pearl2023} in the Dark Energy Spectroscopic Instrument (DESI) One-Percent Survey (\citealt{DESI}). Taken together, these recent studies suggest that galaxy assembly bias is apparent in the spatial distribution of galaxies in the nearby Universe.

Despite the mounting observational evidence in support of galaxy assembly bias, the connection between secondary halo properties and galaxy properties (e.g. stellar mass or star-formation rate; \citealt{kauffmann2015}) remains unconstrained due to challenges affecting observational studies of the galaxy-halo connection. For instance, signatures of one-halo conformity, i.e., the similarity in the observable properties of galaxies within a halo (\citealt{weinmann2006}), can be interpreted as evidence of galaxy assembly bias. However, because these signatures can also be explained by the co-evolution between central and satellite galaxies (e.g. \citealt{pasquali2010,peng2012,wetzel2013,cortese2019,gallazzi2020,oyarzun2023}) and by systematic errors (e.g. \citealt{calderon2018}), using one-halo conformity as a probe for galaxy assembly bias can be difficult. While other work (e.g. \citealt{kauffmann2013}) have attempted to mitigate these degeneracies by searching for signatures of galaxy assembly bias in two-halo conformity instead (the similarity between the physical properties of galaxies in neighboring halos; e.g. \citealt{lacerna2022}), biases in the isolation criterion emerge as a major concern (\citealt{sin2017,tinker2018}).

Instead of studying the impact of galaxy assembly bias on galaxy formation through galactic conformity, other work have turned to the clustering of galaxies. Studies have reported variations in the clustering signal with galaxy environment (\citealt{lehmann2017,zentner2019}) and/or galaxy color (\citealt{montero-dorta2017}). However, the reliability of these results has been questioned in light of systematic errors affecting the photometric group catalogs used for estimating halo masses (\citealt{campbell2015,lin2016}).

With systematic errors hampering observational studies of the galaxy-halo connection, recent efforts have focused on mitigating some of these shortcomings. For instance, the group catalog by \citet{tinker2021,tinker2022} is calibrated with observations of color-dependent galaxy clustering and estimates of the total satellite luminosity, which allows for a more accurate reproduction of the central galaxy quenched fraction. In addition, \citet{tinker2021} used deep photometry from the DESI Legacy Imaging Survey (DLIS; \citealt{dey2019}), allowing for precise accounting of galaxy M$_*$ and total group luminosity (L$_{\mathrm{group}}$). 

For instance, in \citet{oyarzun2022} we exploited the \citet{tinker2021} catalog to estimate the M$_\mathrm{H}$ for over 2000 central galaxies in the SDSS-IV MaNGA survey (Mapping Nearby Galaxies at Apache Point Observatory; \citealt{bundy2015}) and identified variations in the stellar populations of central galaxies at fixed M$_\mathrm{H}$. Specifically, central galaxies with high M$_*$ for their M$_\mathrm{H}$ feature old and Mg-enhanced stellar components. These results, along with equivalent trends recovered in spatially unresolved spectroscopy from the SDSS (\citealt{scholz-diaz2022,scholz-diaz2023,scholz-diaz2024}), are an indication that variations in the assembly histories of central galaxies cannot be fully ascribed to variations in M$_\mathrm{H}$, thus hinting that halo properties other than M$_\mathrm{H}$ may play a role in central galaxy formation. 

In this paper, we test whether these variations in the observable properties of central galaxies at fixed M$_\mathrm{H}$ can be attributed to galaxy assembly bias. To this end, we quantify how halo formation time varies within the SHMR for $\sim$ 150,000 red central galaxies from the SDSS \textsc{main} Galaxy Survey (MGS; \citealt{york2000,gunn2006,dr12}) by utilizing the number density of galaxies within a radius of $R = 10~h^{-1}$~Mpc (\citealt{alpaslan-tinker2020}). The association between halo formation time and the number density of galaxies is motivated by how higher density regions of the large scale structure form at earlier cosmic epochs (e.g. \citealt{hahn2007,lee2017}). 

This manuscript is organized as follows. We present our dataset in Section \ref{2}, methodology in Section \ref{3}, and results in Section \ref{4}. We discuss the implications of our work in Section \ref{5} and summarize in Section \ref{6}. This paper makes use of the spectroscopic M$_*$ measured by \citet{chen2012} and the M$_\mathrm{H}$ estimated by \citet{tinker2021}. The M$_*$ adopt a \citet{kroupa2001} IMF and $h=0.7$. All magnitudes are reported in the AB system (\citealt{oke1983}).

\newpage

\section{Dataset}
\label{2}

\subsection{Galaxy sample}
\label{2.1}

To study the SHMR for central galaxies in the nearby Universe, we turned to the largest sample of nearby galaxies publicly available, the SDSS \textsc{main} Galaxy Survey sample (MGS; \citealt{york2000,gunn2006,dr12}). In particular, we used the \texttt{dr72bright34} subsample of the MGS, which was defined in the construction of the NYU Value-Added Galaxy Catalog (NYU-VAC; \citealt{blanton2005}). Spectroscopic and photometric data outputs are available as part of the eighth data release (DR8) of the SDSS (\citealt{brinchmann2004}). To limit biases in the survey selection function, we restricted our sample of central galaxies to the redshift range $0.04 < z < 0.16$, yielding 434,811 galaxies. To identify central galaxies in the MGS, we turned to the halo-based galaxy group catalog for SDSS by \citet{tinker2021}. We selected all galaxies with satellite probabilities $P_{sat}<0.1$, resulting in 299,409 central galaxies. Further details on the group finder algorithm that we used to identify central galaxies are presented in Section \ref{2.2}. 

Because group catalogs can imprint systematic errors into the M$_\mathrm{H}$ distributions of red and blue galaxies, separating the two populations is recommended (\citealt{lin2016}). To separate red and blue galaxies in the MGS, we followed the Gaussian-mixture model employed by \citet{tinker2021}. This model operates in the space composed of the 4000~\AA\ break index (D$_n4000$) and galaxy luminosity (L$_{\mbox{\scriptsize gal}}$). The outputs of the model were fitted, yielding a threshold of
\begin{equation}
\begin{split}
& \mbox{D$_n^{\mbox{\scriptsize crit}}4000$}(\mbox{L$_{\mbox{\scriptsize gal}}$}) = \\*
 1.42 + \frac{0.35}{2} & \left[1+\texttt{erf}\left(\frac{\log{\mbox{L$_{\mbox{\scriptsize gal}}$}}-9.9}{0.8}\right)\right] 
\end{split}
\end{equation}
to separate the two populations. Here, \texttt{erf(x)} is the error function. We refer the reader to \citet{tinker2021} for more details on how the threshold D$_n^{\mbox{\scriptsize crit}}4000(\mbox{L$_{\mbox{\scriptsize gal}}$})$ was fitted to SDSS data.

This characterization of D$_n^{\mbox{\scriptsize crit}}4000(\mbox{L$_{\mbox{\scriptsize gal}}$})$ was used to identify and remove all blue central galaxies. The resulting sample contains 142,292 red central galaxies, which constitutes the working sample of this study. A sample of red central galaxies has two benefits over a sample of blue central galaxies. First, the majority of centrals are red at the high halo mass end (e.g. \citealt{yang2007}), which is where the halo masses are the most reliable (e.g. \citealt{tinker2022}). Second, our prior characterization of the assembly histories of centrals within the SHMR focused on a passive sample (\citealt{oyarzun2022}), which better corresponds with a red sample.

\begin{figure}
\centering
\includegraphics[width=3.35in]{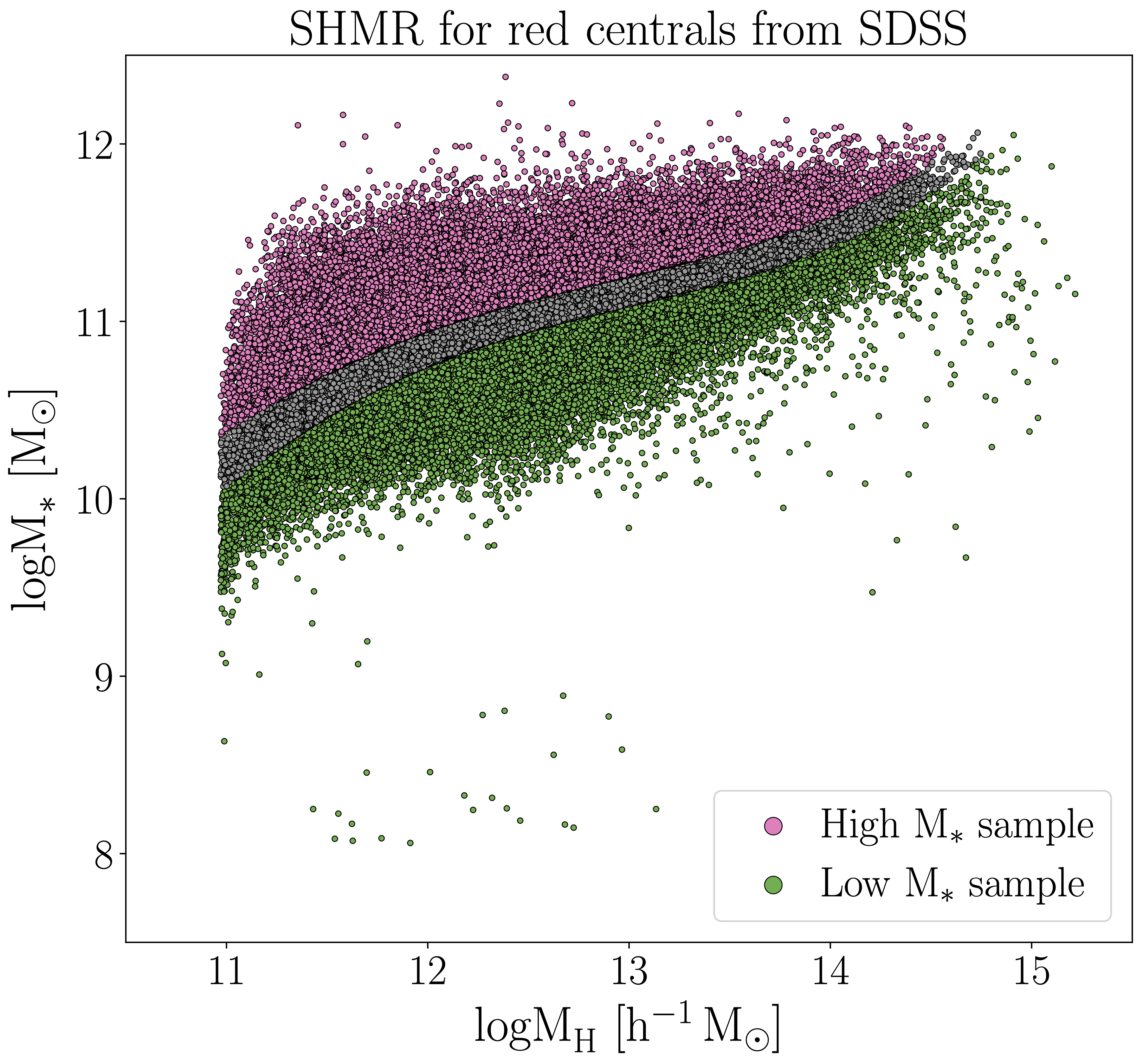}
\caption{Stellar-to-halo mass relation for red central galaxies from SDSS. The M$_*$ were measured in spectra from the SDSS by \citet{chen2012} and the M$_\mathrm{H}$ were estimated with the group-finding algorithm by \citet{tinker2022} on deep photometry from the DLIS (\citealt{tinker2021}). We used the 33rd and 66th percentiles in M$_*$-to-M$_\mathrm{H}$ ratio to divide centrals into high M$_*$ (magenta) and low M$_*$ (green) subsamples.}
\label{fig0}
\end{figure}

\begin{figure*}
\centering
\includegraphics[width=7.1in]{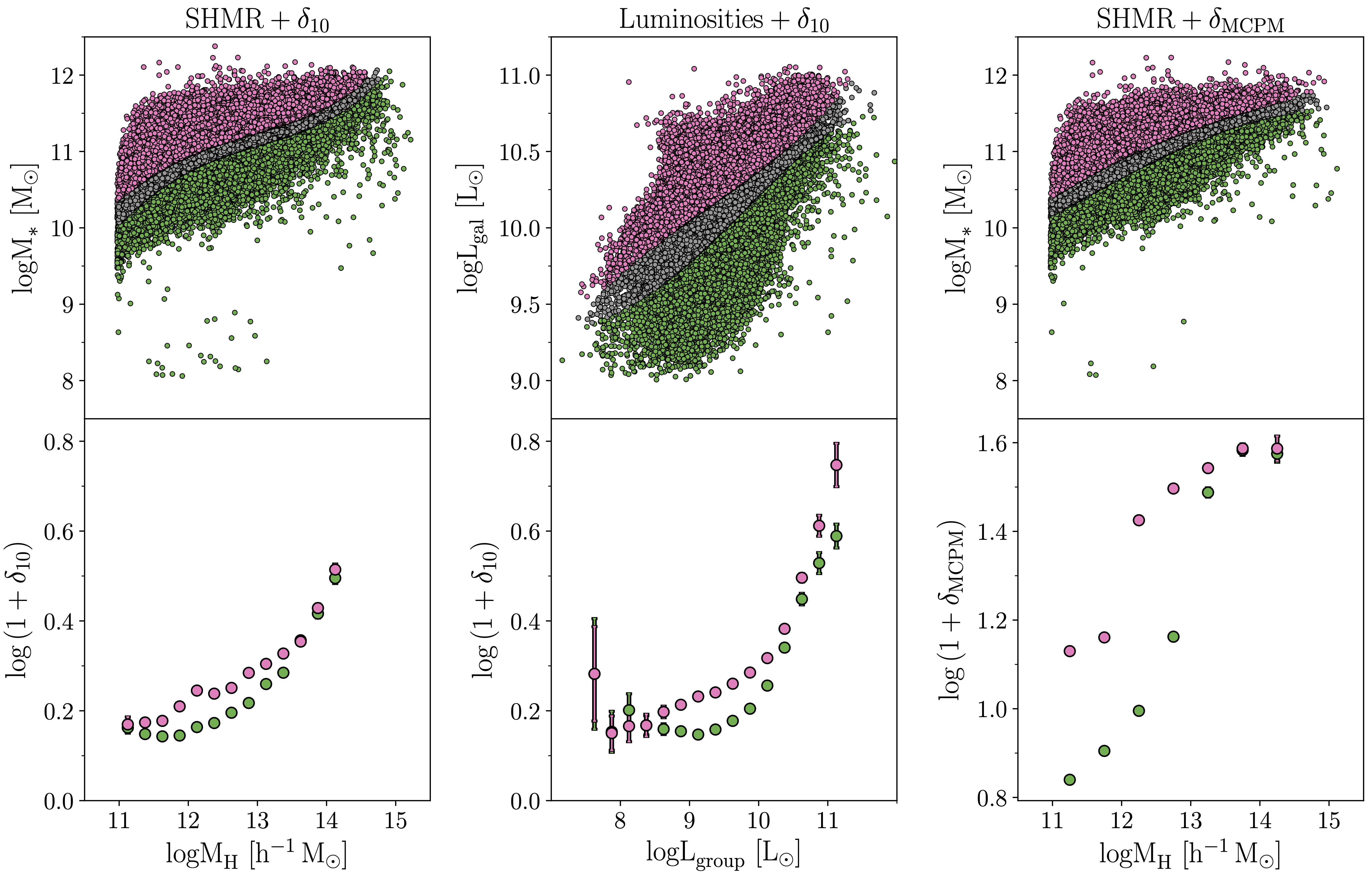}
\caption{\textbf{Left:} The top panel shows the stellar-to-halo mass relation for red central galaxies from SDSS (same as Figure \ref{fig0}). The bottom panel shows the number density of galaxies within a radius $R = 10~h^{-1}$~Mpc as a function of halo mass. High M$_*$ centrals have higher galaxy number densities than low M$_*$ centrals. \textbf{Middle:} Same as the left panel but with galaxy r-band luminosity instead of stellar mass and total group r-band luminosity instead of halo mass. \textbf{Right:} Same as the left panel, albeit with galaxy number density substituted with a metric of the density field measured with the MCPCM Slime Mold reconstruction of the Cosmic Web (\citealt{burchett2020}). Apparent in all three panels is how galaxy stellar mass (or luminosity) varies with proxies for the density of the large scale structure at fixed halo mass (or at fixed group luminosity).}
\label{fig1}
\end{figure*}	

\subsection{Halo masses}
\label{2.2}	

For the identification of central galaxies and to obtain estimates of their M$_\mathrm{H}$, we utilized the halo-based galaxy group finder by \citet{tinker2022}. In this algorithm, the probability of a galaxy being a central depends on both galaxy type and luminosity, enabling accurate reproduction of the central galaxy quenched fraction. After central galaxies and their corresponding groups are determined, halo masses are assigned through abundance matching with the Bolshoi-Planck simulation (\citealt{klypin2016}). At this point, color-dependent galaxy clustering and total satellite luminosity are utilized as observational constraints until the best fitting model is found.

For our analysis, we worked with the publicly available implementation of the \citet{tinker2022} algorithm on SDSS\footnote{\href{https://galaxygroupfinder.net}{https://galaxygroupfinder.net}} (\citealt{tinker2021}). To construct the catalog, \citet{tinker2021} used deep photometry from the DESI Legacy Imaging Survey (DLIS; \citealt{dey2019}), allowing for more accurate accounting of galaxy M$_*$ and group L$_*$ than in other group catalogs. The M$_\mathrm{H}$ for our sample of red central galaxies is shown in Figure \ref{fig0}.

\newpage

\subsection{Stellar masses}
\label{2.3}	

The M$_*$ were retrieved from the SDSS DR8 outputs. They were measured by \citet{chen2012} as part of a PCA-based approach to constrain the star-formation histories and stellar populations for galaxies from SDSS DR7 and BOSS (\citealt{aihara2011}). The small scatter at fixed halo mass shown by these stellar masses ($\sim0.18$~dex) make them ideal for this work (\citealt{tinker2017}). To measure their M$_*$, \citet{chen2012} modeled the optical wavelength range between 3700~\AA\ and 5500~\AA\ of every spectrum with a linear combination of seven eigenspectra. The model space was constructed by generating a spectral library spanning a wide variety of star-formation histories. In this work, we retrieved the M$_*$ that used the \citet{maraston-stromback2011} stellar population synthesis models and \citet{kroupa2001} IMF for the construction of the library.

To determine whether some of the scatter in the red SHMR is driven by secondary halo properties (e.g. \citealt{zentner2014,xu2020}), we defined two subsamples of red central galaxies at fixed M$_\mathrm{H}$: high-M$_*$ and low-M$_*$ centrals. These samples were computed as the 33rd and 66th percentiles in M$_*$-to-M$_\mathrm{H}$ ratio, in similar fashion to \citet{oyarzun2022}. In equation form,
\begin{equation}
\begin{split}
\mbox{High-M$_*$ sample:  }& \frac{\mbox{M$_*$}}{\mbox{M$_\mathrm{H}$}}(\mbox{M$_\mathrm{H}$})>\left[ \frac{\mbox{M$_*$}}{\mbox{M$_\mathrm{H}$}}(\mbox{M$_\mathrm{H}$}) \right]_{.66} \\*
\mbox{Low-M$_*$ sample:  }& \frac{\mbox{M$_*$}}{\mbox{M$_\mathrm{H}$}}(\mbox{M$_\mathrm{H}$})<\left[ \frac{\mbox{M$_*$}}{\mbox{M$_\mathrm{H}$}}(\mbox{M$_\mathrm{H}$}) \right]_{.33}.
\end{split}
\end{equation}

The two subsamples are plotted in Figure \ref{fig0}. High-M$_*$ centrals are plotted in magenta and low-M$_*$ centrals are plotted in green.

\begin{figure*}
\centering
\includegraphics[width=7.1in]{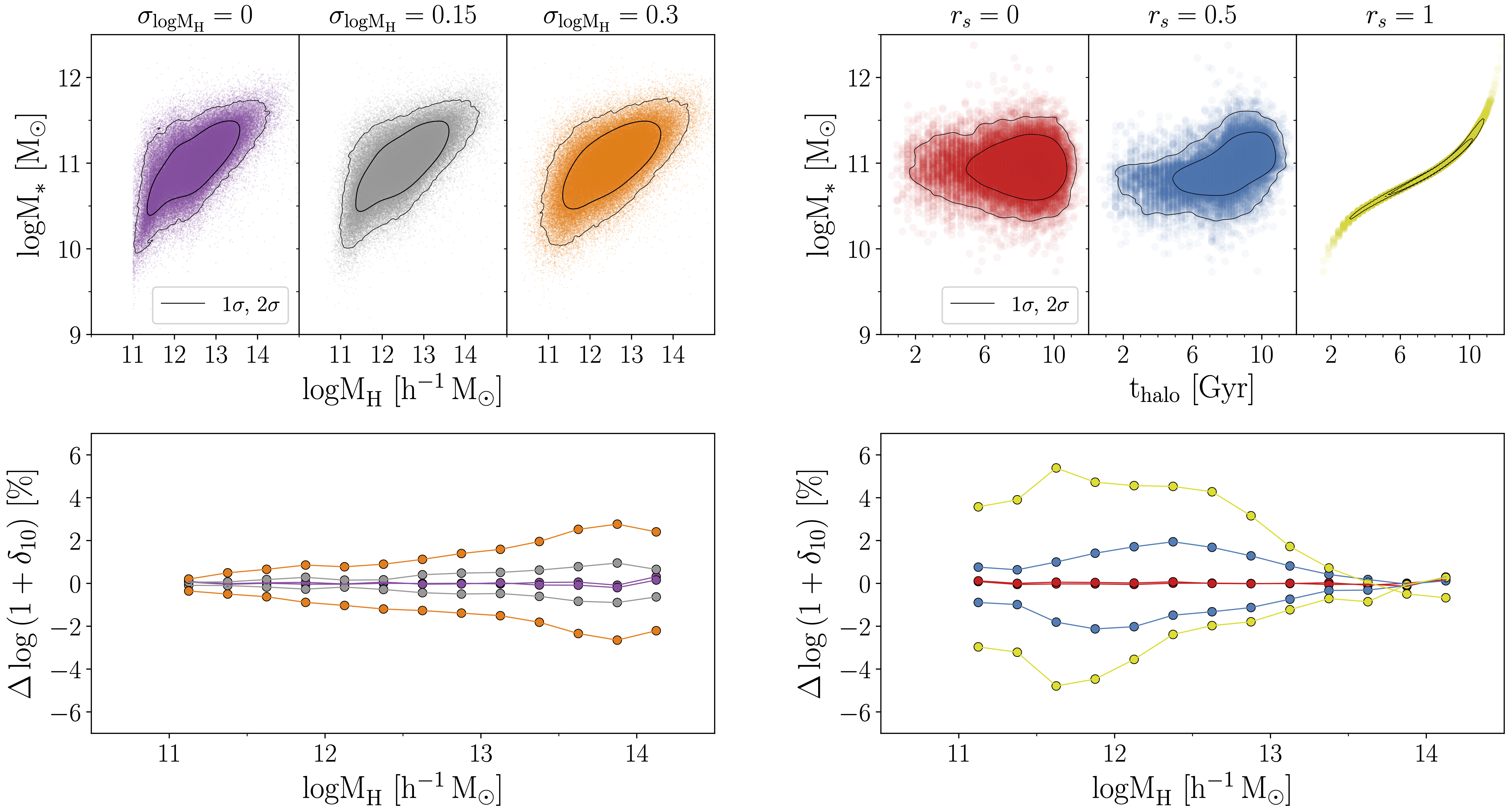}
\caption{Model for interpreting variations in the number density of galaxies with M$_*$ at fixed M$_{\mathrm{H}}$. The free parameters of the model are the random error in the halo mass estimates from the group catalog ($\sigma_{\log{\mathrm{M}_\mathrm{H}}}$) and the Spearman's rank correlation coefficient ($r_s$) between $\mathrm{t}_{\mathrm{halo}}$ and M$_*$ at fixed M$_\mathrm{H}$ (i.e., the galaxy assembly bias signal). \textbf{Left:} The SHMR for red central galaxies from SDSS for different values of $\sigma_{\log{\mathrm{M}_\mathrm{H}}}$ (top). Differences in $\Delta \log(1+\delta_{10})$ between high-M$_*$ and low-M$_*$ centrals for the different values of $\sigma_{\log{\mathrm{M}_\mathrm{H}}}$ (bottom). \textbf{Right:} The correlation between $\mathrm{t}_{\mathrm{halo}}$ and M$_*$ for different values of $r_s$ in the M$_\mathrm{H} = 10^{12.25} - 10^{12.5}~h^{-1}$~M$_{\odot}$ bin (top). Variations in $\Delta \log(1+\delta_{10})$ between high-M$_*$ and low-M$_*$ centrals for the different values of $r_s$ (bottom). We note how $\sigma_{\log{\mathrm{M}_\mathrm{H}}}$ and $r_s$ are only partially degenerate, allowing us to simultaneously quantify their impact on $\Delta \log(1+\delta_{10})$.}
\label{fig2}
\end{figure*}	

\section{Methodology}
\label{3}

\subsection{Galaxy number densities and halo formation times}
\label{3.1}	

In $\Lambda$CDM, higher density regions of the large scale structure collapse at earlier cosmic epochs (\citealt{lemson-kauffmann1999}). We can take advantage of this property of $\Lambda$CDM to use the normalized number density of galaxies within a large radius $R$ as our observational proxy for halo formation time.

In principle, any radius larger than the sizes of halos (i.e., $R\gtrsim 1~h^{-1}$~Mpc) should suffice. However, scales $R\lesssim 5~h^{-1}$~Mpc are not ideal because splashback subhalos can magnify the halo assembly bias signal (e.g. \citealt{sunayama2016}). Moreover, our ability to decouple the effects of M$_\mathrm{H}$ and halo formation time on central galaxy M$_*$ improves as $R$ increases. For these reasons, $R \gtrsim 10~h^{-1}$~Mpc are preferred (\citealt{mansfield2020}). In this work, we opted for $R = 10~h^{-1}$~Mpc to keep the computation time manageable.

The number density of galaxies within $R = 10~h^{-1}$~Mpc (hereafter $\delta_{10}$) was computed for every central as
\begin{equation}
\delta_{10}= \frac{\rho_{z}}{\langle \rho_{z} \rangle}-1, 
\end{equation}
where $\rho_{z}$ is the number density of galaxies around a top-hat sphere of radius $R = 10~h^{-1}$~Mpc and $\langle \rho_{z} \rangle$ is the average $\rho_{z}$ at the redshift of the central. These two quantities were estimated in simulations that randomized the spatial distribution of galaxies, and they account for incompleteness and for edge effects. Further details on how $\delta_{10}$ was computed can be found in \citet{alpaslan-tinker2020}. The average value of $\delta_{10}$ as a function of M$_\mathrm{H}$ for our red central galaxies is shown in Figure \ref{fig1}.

To connect galaxy number density and halo formation time, we turned to the MultiDark Planck 2 Simulation (MDPL2; \citealt{prada2012}). We first measured M$_\mathrm{H}$, $\delta_{10}$, and $\mathrm{t}_{\mathrm{halo}}$ for every host halo in the simulation, where $\mathrm{t}_{\mathrm{halo}}$ is the lookback time at which the host halo assembled half of its total mass at $z=0$ (e.g. \citealt{tojeiro2017}). Although all halo formation times reported in this manuscript were obtained according to the aforementioned definition, the main conclusions of this paper also stand if we instead use $\mathrm{t}_{\mathrm{halo}}^{11.5}$, which corresponds to the lookback time at which a critical halo mass of M$_\mathrm{H} =  10^{11.5}~h^{-1}$~M$_{\odot}$ was reached (more in Section \ref{4}).

For the measurement of the number density of galaxies in MDPL2, only subhalos with $V_{\mathrm{peak}}>100$~km\,s$^{-1}$ were counted, where $V_{\mathrm{peak}}$ is the maximum peak circular rotation velocity across all redshifts. This choice can be interpreted as counting only subhalos with M$_\mathrm{H}\gtrsim 10^{11}~h^{-1}$~M$_{\odot}$ (\citealt{kravtsov2004,zehavi2019}). Then, we ensured that the dynamic range of $\delta_{10}$ in the data and in the simulation were comparable. To this end, we implemented an abundance matching approach, i.e., the halos were rank ordered in $\delta_{10}$ at fixed M$_\mathrm{H}$ and their $\delta_{10}$ values were matched.

\subsection{Galaxy assembly bias model}
\label{3.2}

Here, we describe the analytical model we used to associate differences in $\delta_{10}$ at fixed M$_\mathrm{H}$ with variations in halo formation time. Our model has two parameters: the error in the M$_\mathrm{H}$ reported by the group catalog and the strength of the correlation between halo formation time and stellar mass at fixed M$_\mathrm{H}$.

Imprecise halo mass estimates can broaden the observed scatter of the SHMR, potentially biasing our high-M$_*$ and low-M$_*$ subsamples (see Figure \ref{fig0}). To account for this effect, we added random perturbations to the measured halo masses prior to the sample assignment step (Section \ref{2.3}). These random perturbations were drawn from a normal distribution characterized by a scatter with magnitude `$\sigma_{\log{\mathrm{M}_\mathrm{H}}}$'. The impact that different values of $\sigma_{\log{\mathrm{M}_\mathrm{H}}}$ have on the galaxy assembly bias signal is shown in the left panels of Figure \ref{fig2}. It is apparent in this figure that errors in the group catalog broaden the scatter of the SHMR and induce variations in $\delta_{10}$ toward M$_\mathrm{H}>10^{12.5}~h^{-1}$~M$_{\odot}$.

The second free parameter in our model is the strength of the galaxy assembly bias signal at a given M$_\mathrm{H}$, which can be quantitatively expressed via the Spearman's rank correlation coefficient ($r_s$). In the case of halo formation time and galaxy stellar mass, $r_s$ can be written as:
\begin{equation}
r_s(\mbox{M}_h) = 1 - \frac{6 \sum{ [r(\mbox{M}_*(\mbox{M}_h)) - r(\mathrm{t}_\mathrm{halo}(\mbox{M}_h)) ] ^ 2}}{n(\mbox{M}_h)(n^2(\mbox{M}_h)-1)}.
\end{equation}
Here, $r(X)$ symbolizes the ordered rank of the galaxies in variable $X$, which in our case takes the values $X=$~M$_*$ and $X=\mathrm{t}_{\mathrm{halo}}$ at the corresponding M$_\mathrm{H}$ bin. The total number of galaxies in the M$_\mathrm{H}$ bin is represented by `$n(\mbox{M}_\mathrm{H})$'. If $r_s = 1$, then M$_*$ and $\mathrm{t}_{\mathrm{halo}}$ are perfectly correlated (i.e., the perfect galaxy assembly bias case). In the case of no galaxy assembly bias, $r_s = 0$. Values in the range  $r_s = (0,1)$ correspond to different strengths of the galaxy assembly bias signal, as illustrated in the right panels of Figure \ref{fig2}.

In the context of our work, we had at our disposal a collection of host halos with ($\mathrm{t}_{\mathrm{halo}}$, $\delta_{10}$) measured in MDPL2 and a collection of galaxies with (M$_*$, $\delta_{10}$) measured in SDSS. The correspondence between M$_*$ and $\mathrm{t}_{\mathrm{halo}}$ at fixed M$_\mathrm{H}$ is determined by the value of $r_s$, which in turn dictates how host halos and galaxies are mapped onto each other. The mapping between host halos and galaxies was performed in narrow M$_\mathrm{H}$ bins, with a unique value for $r_s$ at all M$_\mathrm{H}$ (i.e., we assume $r_s$ not to vary with M$_\mathrm{H}$). We note that blue central galaxies were included in this process to ensure appropriate matching between the populations of halos and galaxies.

\subsection{Model fitting}
\label{3.3}

We used our two-dimensional ($\sigma_{\log{\mathrm{M}_\mathrm{H}}}$, $r_s$) model to fit the observed variations in $\log(1+\delta_{10})$ at fixed M$_\mathrm{H}$. These variations --- hereafter $\Delta \log(1+\delta_{10})$ --- were computed relative to the median value for $\log(1+\delta_{10})$ in a given M$_\mathrm{H}^{i}$ bin.
\begin{equation}
\begin{split}
\mbox{High-M$_*$ sample:  } &\Delta \log(1+\delta_{10}(\mbox{high-M}_*, \mbox{M}_\mathrm{H}^{i})) = \nonumber \\*
&\log(1+\delta_{10}(\mbox{high-M}_*, \mbox{M}_\mathrm{H}^{i}))\, - \nonumber \\* &\mathrm{Median}[\log(1+\delta_{10}(\mbox{M}_*, \mbox{M}_\mathrm{H}^{i}))].
\end{split}
\end{equation}
\begin{equation}
\begin{split}
\mbox{Low-M$_*$ sample:  } &\Delta \log(1+\delta_{10}(\mbox{low-M}_*, \mbox{M}_\mathrm{H}^{i})) = \nonumber \\*
&\log(1+\delta_{10}(\mbox{low-M}_*, \mbox{M}_\mathrm{H}^{i}))\, - \nonumber \\* &\mathrm{Median}[\log(1+\delta_{10}(\mbox{M}_*, \mbox{M}_\mathrm{H}^{i}))].
\end{split}
\end{equation}

The uncertainties on
\begin{equation}
\begin{split}
\Delta \log(1+\delta_{10}&(\mbox{high-M}_*, \mbox{M}_\mathrm{H}^{i})) \\* &\mbox{and} \\*  
\Delta \log(1+\delta_{10}&(\mbox{low-M}_*, \mbox{M}_\mathrm{H}^{i}))
\end{split}
\end{equation}
were propagated from the uncertainties on
\begin{equation}
\delta_{10}(\mbox{high-M}_*, \mbox{M}_\mathrm{H}^{i}) \mbox{ and } \delta_{10}(\mbox{low-M}_*, \mbox{M}_\mathrm{H}^{i}),
\end{equation}
which were computed by bootstrapping the galaxies selected for the computation of $\delta_{10}$. The errors on the average M$_\mathrm{H}$ of every bin were not fitted for, and the different M$_\mathrm{H}$ bins were assumed independent.

This set of $\Delta \log(1+\delta_{10})$ measurements as a function of M$_{\mathrm{H}}$ compose the dataset that was fitted with the galaxy assembly bias model. Explicitly, the quantities that we fitted were
\begin{equation}
\begin{split}
\Delta \log(1+\delta_{10}&(\mbox{high-M}_*, \mbox{M}_\mathrm{H}^{i})) \\* &\mbox{and} \\*
\Delta \log(1+\delta_{10}&(\mbox{low-M}_*, \mbox{M}_\mathrm{H}^{i})),
\end{split}
\end{equation}
as they appear in the bottom panels of Figure \ref{fig2}. We highlight that $\Delta \log(1+\delta_{10})$ is only explicitly dependent on three quantities: M$_\mathrm{H}$, M$_*$, and $\delta_{10}$. In observational space, these quantities were measured in SDSS. In model space, the M$_\mathrm{H}$ correspond to the M$_\mathrm{H}$ in MDPL2, the M$_*$ were assigned according to the value of $r_s$, and the $\mathrm{t}_{\mathrm{halo}}$ were measured in MDPL2 (see Sections \ref{3.1} and \ref{3.2}). Although the $\delta_{10}$ in model space were measured in MDPL2, they were abundance matched to the values in SDSS (end of Section \ref{3.1}).

In the computation of the posterior, we adopted uniform, bounded priors for both parameters, i.e.,
\begin{equation}
\begin{split}
p(\sigma_{\log{\mathrm{M}_\mathrm{H}}}) &= \mbox{TopHat}\,[0, 1] \\*
p(r_s) &= \mbox{TopHat}\,[0, 1]. 
\end{split}
\end{equation}

The Markov chain Monte Carlo sampler \texttt{emcee} (\citealt{emcee}) was used to constrain the posterior. We implemented a setup of 10,000 walkers, 2400-step chains, and 200 burn-in steps. 

\begin{figure}[t!]
\centering
\includegraphics[width=3.35in]{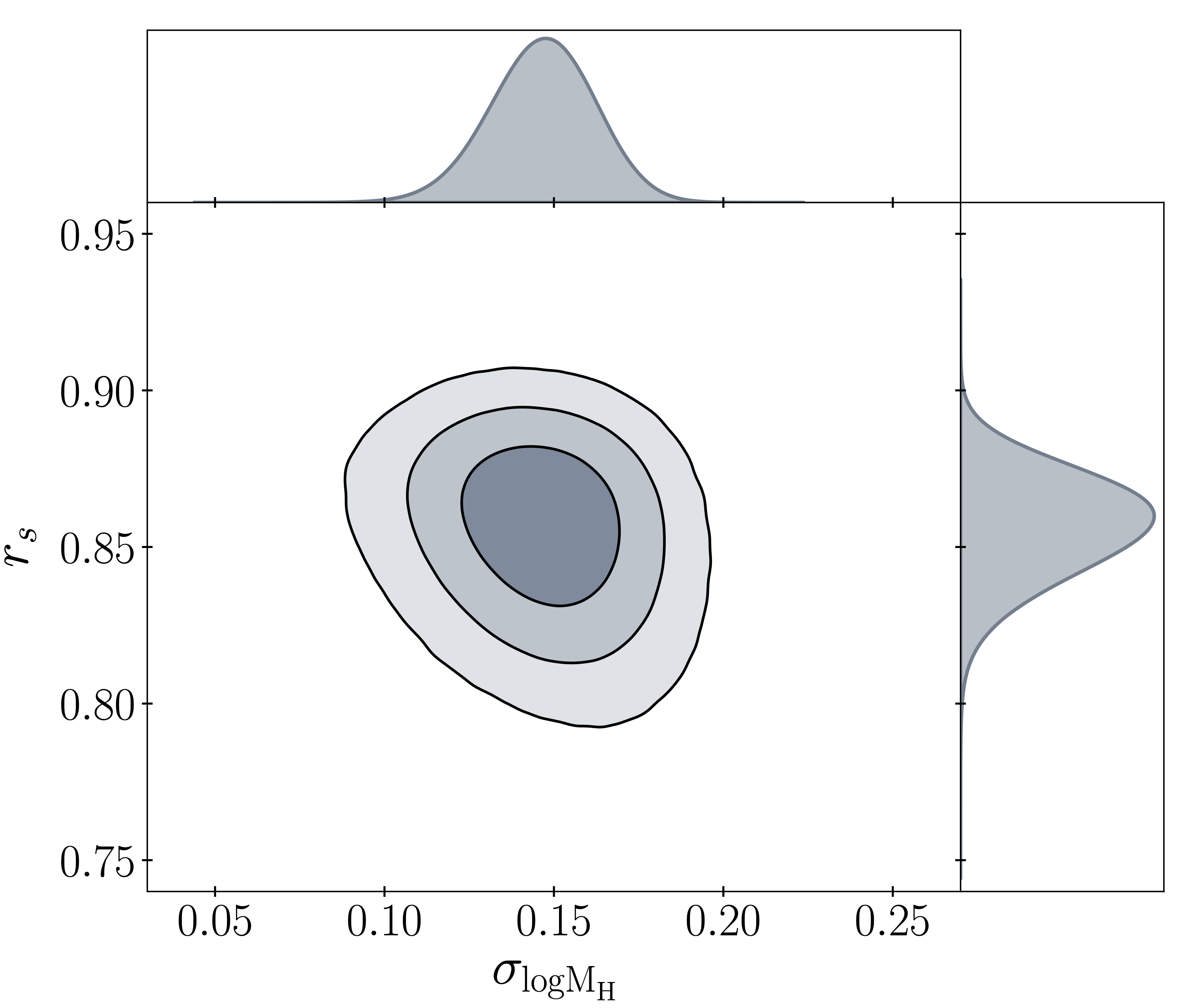}
\caption{Posterior distribution (1$\sigma$, 2$\sigma$, 3$\sigma$ contours) of the model that fits for variations in galaxy number density within the SHMR for red central galaxies. The x-axis shows the random error in the halo mass estimates from the group catalog. The y-axis shows the Spearman's rank correlation coefficient between $\mathrm{t}_{\mathrm{halo}}$ and M$_*$ at fixed M$_\mathrm{H}$ (i.e., the galaxy assembly bias signal). The no-correlation case corresponds to $r_s = 0$ and the perfect correlation case corresponds to $r_s = 1$. The absence of assembly bias ($r_s = 0$) is inconsistent with the data.}
\label{fig3}
\end{figure}

\begin{figure*}
\centering
\includegraphics[width=7.1in]{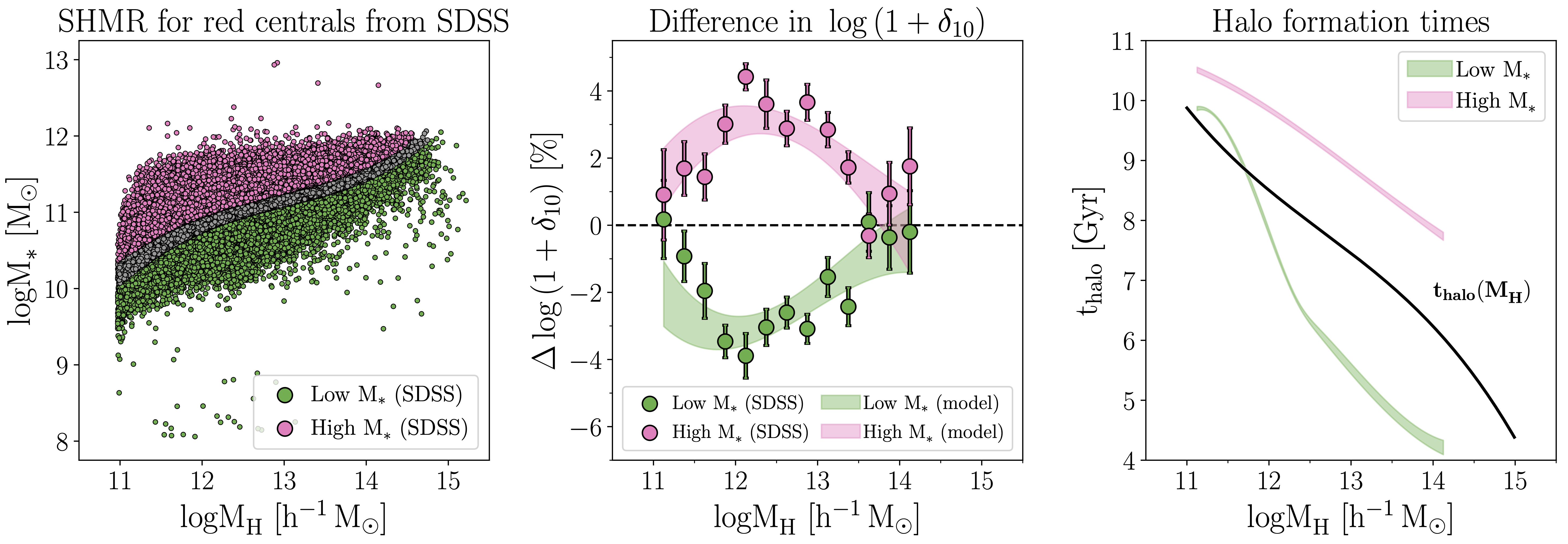}
\caption{Model solution after convergence. \textbf{Left:} The SHMR for red central galaxies from SDSS highlighting the two subsamples (same as Figure \ref{fig0}). High-M$_*$ centrals are plotted in magenta ($>66$th percentile in the SHMR) and low-M$_*$ centrals are plotted in green ($<33$rd percentile in the SHMR). \textbf{Middle:} Difference in galaxy number density within $R = 10~h^{-1}$~Mpc between high- and low-M$_*$ red centrals as a function of M$_{\mathrm{H}}$ (datapoints). The shaded regions show the 1$\sigma$ difference in galaxy number density from a random sampling of the converged posterior. \textbf{Right:} 1$\sigma$ intervals of the $\mathrm{t}_{\mathrm{halo}}$ of high-M$_*$ and low-M$_*$ red centrals as constrained by the model. Also shown in black is the median $\mathrm{t}_{\mathrm{halo}}$ of host halos in the MDPL2 simulation as a function of M$_\mathrm{H}$. Note that the average between the $\mathrm{t}_{\mathrm{halo}}$ of high-M$_*$ and low-M$_*$ red centrals differs from the median $\mathrm{t}_{\mathrm{halo}}$ in MDPL2 because the median $\mathrm{t}_{\mathrm{halo}}$ was computed for all centrals (i.e., both red and blue).}
\label{fig4}
\end{figure*}

\section{Results}
\label{4}

Plotted in the top left panel of Figure \ref{fig1} is the SHMR for red central galaxies from SDSS separated into two bins of M$_*$ at fixed M$_\mathrm{H}$ (Section \ref{3.1}). The bottom left panel shows that centrals with high-M$_*$ have systematically higher galaxy number densities within $R = 10~h^{-1}$~Mpc than centrals with low-M$_*$, especially in the M$_\mathrm{H}$ range between $10^{11.5}-10^{13.5}~h^{-1}$~M$_{\odot}$. This result is evidence that the M$_*$ of galaxies vary with the density of the large scale structure at fixed M$_\mathrm{H}$. 

We can confirm that this result holds true after controlling for all three quantities involved in the measurement. Upon substitution of M$_\mathrm{H}$ with group r-band luminosity and M$_*$ with central galaxy r-band luminosity, differences in $\delta_{10}$ remain significant (middle panel of Figure \ref{fig1}). We also controlled for $\delta_{10}$ by utilizing reconstructions of the Cosmic Web with the Monte Carlo Physarum Machine (MCPM) Slime Mold algorithm (\citealt{burchett2020}). In contrast to the number density of galaxies, which is sensitive to the abundance of luminous matter in the large scale structure, the MCPM Slime Mold catalog (\citealt{wilde2023}) provides estimates for the density field in its totality, including dark-matter (\citealt{hasan2023b}). This was achieved through the adaptation of an algorithm designed to characterize the growth of `slime mold', enabling reconstruction of the tri-dimensional structure of the Cosmic Web. We used the publicly available 
\footnote{\href{https://www.sdss.org/dr18/data_access/value-added-catalogs/?vac_id=99}{https://www.sdss.org/dr18/data\_access/value-added-catalogs/?vac\_id=99}}
measurements in replacement of the number density of galaxies, achieving a consistent --- and perhaps even more significant --- signal in the right panel of Figure \ref{fig1}. Taken together, these results are evidence that the baryonic properties of central galaxies correlate with the density of the large scale structure after controlling for the mass at the halo scale.

While differences in $\delta_{10}$ are suggestive of differences in halo formation time (e.g. \citealt{lee2017,alpaslan-tinker2020}), modeling is required in order to account for other possible sources of variation in $\delta_{10}$. The framework introduced in Section \ref{3.2} serves this purpose by simultaneously modeling the effects of M$_\mathrm{H}$ errors and galaxy assembly bias on $\delta_{10}$ variations within the red SHMR. We fitted this model to the observed differences in $\delta_{10}$ (left panel of Figure \ref{fig1}) and obtained the posterior distribution shown in Figure \ref{fig3}. Inspection of this figure reveals that random errors in the halo mass estimates and a galaxy assembly bias signal are both needed to explain the differences in $\delta_{10}$, with $\sigma_{\log{\mathrm{M}_\mathrm{H}}}\sim 0.15$~dex and $r_s \sim 0.85$ best reproducing the observations. Errors in the M$_\mathrm{H}$ of this magnitude are within expectations from the group finder algorithm ($\sigma_{\log{\mathrm{M}_\mathrm{H}}}\lesssim 0.2$; \citealt{tinker2022}) and the marginal posterior probability for $r_s$ rejects the absence of galaxy assembly bias ($r_s = 0$) with confidence~$>5\sigma$.

The model solutions are plotted along with the data in Figure \ref{fig4}. As revealed by the middle panel, the differences in galaxy number density at fixed M$_\mathrm{H}$ are well reproduced by the converged model. Plotted in the right panel are the the halo formation times of low- and high-M$_*$ red centrals that have been inferred by our model. More massive dark-matter halos form at later times, and halos that are old for their M$_\mathrm{H}$ host galaxies with higher M$_*$. We can confirm that this result also holds true for different definitions of halo formation time: at fixed M$_\mathrm{H}$, halos with older $\mathrm{t}_{\mathrm{halo}}^{11.5}$ (as defined in Section \ref{3.1}) host higher M$_*$ galaxies.

\section{Discussion}
\label{5}

Variations in the baryonic properties of galaxies at fixed M$_{\mathrm{H}}$ have been observed for a few years (e.g. \citealt{oyarzun2022,scholz-diaz2022,scholz-diaz2023,scholz-diaz2024}). However, diagnosing the source of these variations has been difficult due to the stochastic nature of galaxy formation and because of the systematic errors affecting observational halo diagnostics (e.g. \citealt{calderon2018,tinker2018}). In this context, we have determined that some of these variations at fixed M$_{\mathrm{H}}$, in particular differences in stellar mass and/or galaxy luminosity, are correlated with how luminous matter is arranged in large ($R = 10~h^{-1}$~Mpc) scales. In this section, we discuss the biases and/or the underlying physical processes that could be driving this connection.

\subsection{Systematic errors in the group catalog}
\label{5.2}

It is important to address how systematic errors in our approach could fabricate a signal in the galaxy number densities. Errors in the halo mass estimates from the group catalog are particularly relevant, which is the reason for their inclusion in the model. As we show in Figure~\ref{fig3}, random errors in the group catalog alone cannot explain the differences in galaxy number densities at fixed M$_\mathrm{H}$. That said, should the errors in the group finder not be random (i.e., stellar- or halo-mass dependent), then any signal in the galaxy number densities could be fabricated. For instance, weak-lensing analyses have revealed differences between the underlying M$_\mathrm{H}$ distributions of red and blue central galaxies in the \citet{yang2007} group catalog, indicating that the scatter of the SHMR correlates with the physical properties of galaxies in an unaccounted for manner (\citealt{lin2016}).

Our approach accounts for these difficulties in multiple ways. First, our detection is not reliant on differences between red and blue centrals, since the latter were removed from our sample. Second, the impact of the color-dependent spatial distribution of galaxies on halo mass estimation is accounted for, as the \citet{tinker2022} algorithm is designed to fit for the clustering of red and blue galaxies separately. It is also reassuring that the SDSS implementation of the algorithm can reproduce the weak-lensing derived SHMR (\citealt{mandelbaum2016}), especially for red centrals in our M$_\mathrm{H}$ regime (\citealt{tinker2021}). Finally, the fact that we also recover the signal when using total group luminosity instead of M$_\mathrm{H}$ (Figure \ref{fig1}) is an indication that the signal is not fabricated by how the halo masses are assigned by the group finder algorithm. 

\begin{figure}
\centering
\includegraphics[width=3.2in]{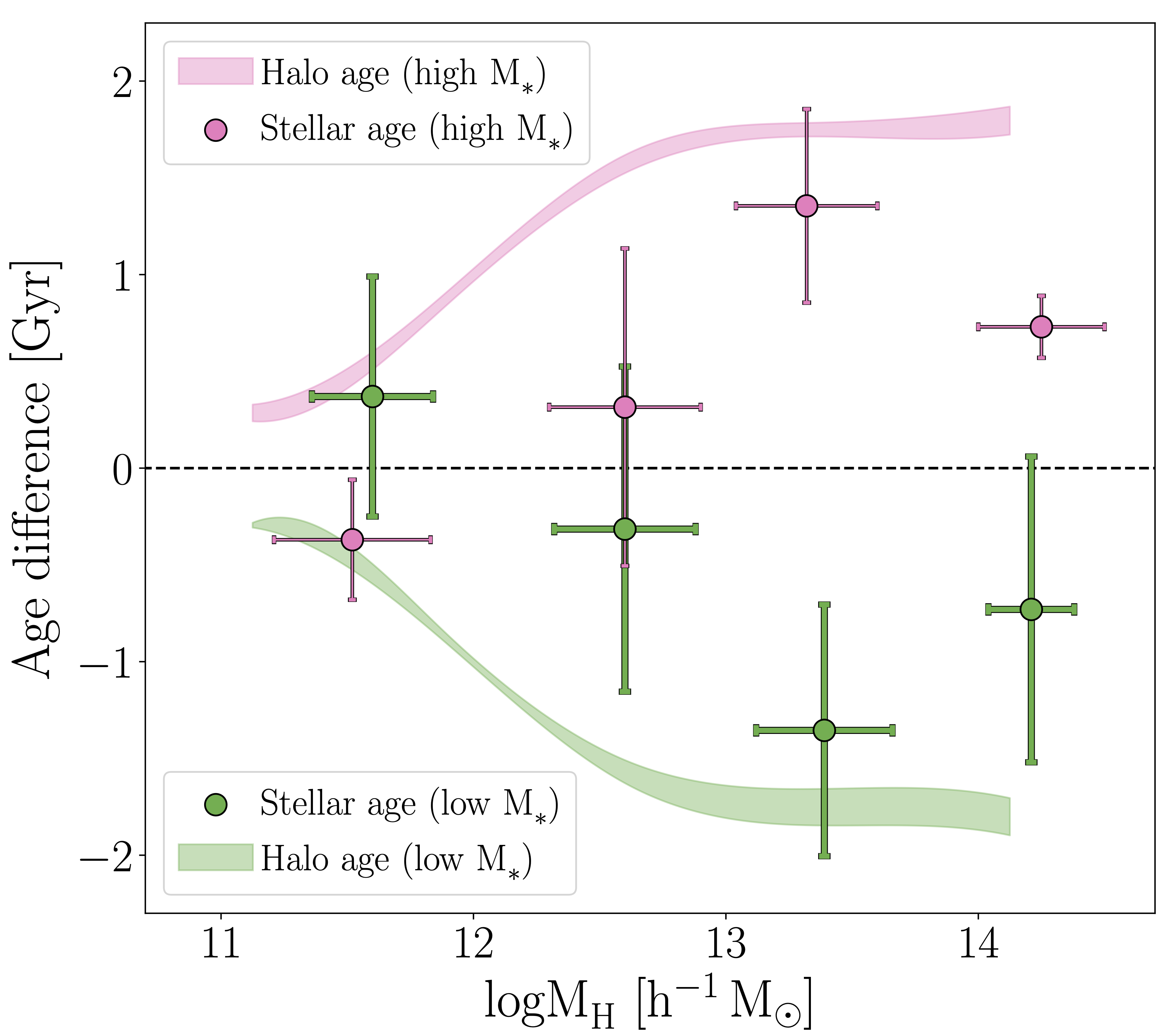}
\caption{Variations in halo (this work) and galaxy (\citealt{oyarzun2022}) formation times within the SHMR for red central galaxies. High- and low-M$_*$ centrals were defined according to the 33rd and 66th percentiles in M$_*$-to-M$_\mathrm{H}$ ratio in the red SHMR (magenta and green, respectively). The differences for both high- and low-M$_*$ centrals are relative to the median formation time at a given M$_\mathrm{H}$. The differences in halo formation time correspond to the 1$\sigma$ spread of a random posterior sampling (shaded regions). The differences in galaxy formation time correspond to the average mass-weighted stellar ages derived from stellar population fitting of MaNGA spectra (datapoints). Differences in the formation times are $\sim 1-3$~Gyr for both halos and galaxies.}
\label{fig5}
\end{figure}

Systematic errors in group catalogs can also stem from central and satellite galaxy mis-classification. \citet{lin2016} showed that satellite contamination can affect the correlation functions and the dark matter profiles of central galaxy samples. Yet, as also pointed out by \citet{lin2016}, satellites that are mis-classified as centrals reflect the large-scale bias of their higher M$_\mathrm{H}$ hosts. As the normalized densities converge at high M$_\mathrm{H}$ (Figure \ref{fig2}), central and satellite galaxy mis-classification would weaken the assembly bias signal; a conclusion that we can confirm applies to our dataset after conducting simulations of galaxy miss-classifications. Moreover, adopting a central selection criterion of P$_{sat}<0.01$ instead of P$_{sat}<0.1$ does not impact our results.

Any systematic errors arising from our M$_*$ are unlikely. At fixed M$_\mathrm{H}$, the measurement error of our stellar masses is $\Delta$M$_* < 0.18$~dex (\citealt{tinker2017}), much smaller than the $\sim 0.4$~dex separating our two M$_*$ subsamples (left panel of Figure \ref{fig0}). Moreover, the signal is also recovered after substituting M$_*$ with L$_*$ (Figure \ref{fig1}). The normalized number densities are similarly robust, with the flux limit of the DLIS reaching $22.8$ mag at 5$\sigma$ (\citealt{dey2019}), which is $5-6$ magnitudes deeper than the $m_*$ of the galaxy luminosity function at $z=0.1$ (\citealt{blanton2005}). We also confirm that the signal is recovered with independent metrics for the density of the large scale structure (i.e., the Slime Mold reconstruction of the Cosmic Web; Figure \ref{fig1}).

Still, quantifying the magnitude and significance of galaxy assembly bias with other probes for M$_\mathrm{H}$ remains a priority. Weak lensing can be used to search for systematic differences between the total mass surface density profiles of high- and low-M$_*$ centrals (e.g. \citealt{lin2016}). In addition, host M$_\mathrm{H}$ can also be estimated through the tight correlation between central M$_*$, M$_\mathrm{H}$, and M$^{max}_*$ (where M$^{max}_*$ is the the stellar mass of the central out to 100 kpc; \citealt{huang2020}) or via the stellar-to-total dynamical mass relation for central galaxies (\citealt{scholz-diaz2024}). Also key will be alternative methods for estimating halo formation times in observations (e.g. machine learning algorithms; \citealt{lyu2024}).

\subsection{Analytic model for galaxy assembly bias}
\label{5.4}

The magnitude of the difference in halo formation time between the two M$_*$-selected subsamples that has been inferred by our method is $\sim 1-3$~Gyr (Figure \ref{fig5}). Differences of this magnitude are consistent with the variations in stellar age that we found in \citet[also plotted in Figure \ref{fig5}]{oyarzun2022} after similarly dissecting the red SHMR of nearby galaxies. In an approach that is analogue to that of Section \ref{3.1}, we also separated central galaxies according to the 33rd and 66th percentiles in M$_*$-to-M$_\mathrm{H}$ ratio, recovering variations in the timescales of galaxy formation at fixed M$_\mathrm{H}$ that are not drastically different from the variations in the timescales of halo formation. In this subsection, we will show that variations in the stellar ages and stellar masses of galaxies with the large-scale environment (at fixed M$_\mathrm{H}$) can be interpreted as a natural outcome of linear structure-formation theory and the regulation of star-formation through energy-conserving feedback. 

We will first discuss the difference in stellar age with galaxy environment in large scales (Figure \ref{fig5}). In this context, $R = 10~h^{-1}$~Mpc is a natural scale to consider because that is the non-linear mass scale in the local Universe. The redshift at which a particular region of the Universe reaches a virialized or collapsed mass fraction (i.e., M$_\mathrm{H}$) depends on the large scale overdensity, with overdense regions having accelerated structure formation. According to extended Press-Schechter theory, the variance in redshift at which a region of the Universe within a radius $R$ will reach a particular collapsed fraction is 
\begin{equation}
\sigma_z = \frac{\sigma(R)}{1.69\,(1+z)}, 
\end{equation}
where $\sigma(R)$ is the variance in the density field (\citealt{barkana-loeb2004,wyithe-loeb2004}). For $R = 10$~cMpc at $z=0$, the variance in redshift is $\sigma_z\approx0.6$. These variations can be linked to differences in the formation of structure for an overdense Universe, i.e.,
\begin{equation}
dt = \left|\frac{dt}{dz} \right| dz = \frac{\sigma_z}{(1+z)\,H(z)} \approx 1~\mbox{Gyr at } z=0. 
\end{equation}
Variations of this magnitude are consistent with the observed differences in age between the stellar populations of galaxies in large scale overdensities and underdensities at fixed M$_\mathrm{H}$ (Figure \ref{fig5}). 
 
With this result in hand, we can predict the differences in M$_*$ between old and young populations at fixed M$_\mathrm{H}$. Stellar populations with ages of $10$~Gyr formed before quasars were dominant in regulating stellar growth (e.g. \citealt{croton2006a,croton2006b}). Under the ansatz that stellar growth is set by the balance between the binding energy of the star-forming gas and the mechanical energy from supernovae (the base assumption in almost all semi-analytic models and numerical simulations), the ratio between stellar and halo mass corresponds to
\begin{equation}  
\frac{M_*}{\mathrm{M}_\mathrm{H}} \propto \mathrm{M}_\mathrm{H}^{2/3} (1+z),
\end{equation}
as in \citet{dekel-woo2003} and \citet{wyithe-loeb2003}. This implies that the range in M$_*$ at fixed M$_\mathrm{H}$ owing to variations in the formation time is
\begin{equation}
\Delta \mbox{M}_* = \frac{(1+z+\sigma_z)}{(1+z-\sigma_z)}. 
\end{equation}
For the derived value of $\sigma_z\approx0.6$ for $R = 10$~cMpc, $\Delta \mbox{M}_*\approx 3$ for stars forming at $z=0$ and $\Delta \mbox{M}_*\approx 2$  for stars forming at $z=1$. These values match the range of stellar mass that separates the high-M$_*$ and low-M$_*$ subsamples ($\Delta \mbox{logM}_*\approx 0.4$; Figure \ref{fig0}).    

\subsection{Galaxy assembly bias in simulations}
\label{5.1}

The results shown in Figure \ref{fig4} and the analytical arguments presented in Section \ref{5.4} can be interpreted as an underlying correlation between M$_*$ and halo formation time. In this scenario, the properties of galaxies are dependent not only on the masses of dark matter halos, but also on secondary halo properties like formation time or assembly history. A fixed M$_{\mathrm{H}}$, older halos form deeper gravitational potentials that facilitate the assembly of higher M$_*$ central galaxies (e.g. \citealt{matthee2017}).

Our model associates variations in $\delta_{10}$ across  M$_\mathrm{H} \sim 10^{11}-10^{14}~h^{-1}$~M$_{\odot}$ with a strong correlation between halo formation time and stellar mass ($r_s\sim 0.85$) across this M$_\mathrm{H}$ range. In broad agreement with this result, the correlation between halo age and M$_*$ is present across the M$_\mathrm{H}\sim10^{11}-10^{13.5}~h^{-1}$~M$_{\odot}$ range in the IllustrisTNG simulation (\citealt{zehavi2018,xu2020}). Interestingly, the galaxy assembly bias in IllustrisTNG is also present when using the number density of particles within $R\sim 5~h^{-1}$ Mpc instead of $\mathrm{t}_{\mathrm{halo}}$ (\citealt{zehavi2018}), which is consistent with our analysis in SDSS (albeit with the considerations regarding the choice of $R$ outlined in Section \ref{3.1}).

We note that the magnitude of the $\delta_{10}$ differences peak around M$_\mathrm{H} \sim 10^{12}$~M$_{\odot}$. However, because we adopt a constant value for $r_s$ across all M$_\mathrm{H}$, we may not interpret any dependence of the $\delta_{10}$ variations on M$_\mathrm{H}$ as a signature of a M$_\mathrm{H}$-dependent galaxy assembly bias signal. Some theoretical work --- e.g. IllustrisTNG (\citealt{vogelsberger2014a,nelson2015}) --- predict the galaxy assembly bias signal to decrease in magnitude toward M$_\mathrm{H} > 10^{13.5}~h^{-1}$~M$_{\odot}$ (\citealt{zehavi2018,xu2020}). On the other hand, galaxy assembly bias remains strong in the UniverseMachine (\citealt{behroozi2019}) for M$_\mathrm{H}>10^{14}~h^{-1}$~M$_{\odot}$ (\citealt{bradshaw2020}). The EAGLE simulation (\citealt{crain2015,schaye2015}) is at the other end of the spectrum, predicting that halo formation time has little impact above M$_\mathrm{H}\gtrsim 10^{11.5}~h^{-1}$~M$_{\odot}$ (\citealt{matthee2017}).

Some of the discrepancy between these simulations may originate in how halo and galaxy mergers are handled, especially given that mergers become increasingly important as M$_\mathrm{H}$ increases (\citealt{huang2013a,huang2013b,huang2018,huang2020,oyarzun2019,angeloudi2024}). In the case of our work, this is apparent in the little sensitivity of $\delta_{10}$ variations to the correlation between $\mathrm{t}_{\mathrm{halo}}$ and M$_*$ toward M$_\mathrm{H}>10^{13.5}~h^{-1}$~M$_{\odot}$ (Figure \ref{fig3}). This justifies the choice of fitting for a single, M$_\mathrm{H}$-independent value for $r_s$, and it ultimately stems from how $\mathrm{t}_{\mathrm{halo}}$ and $\delta_{10}$ are correlated at these M$_\mathrm{H}$ in MDPL2.

For a more quantitative comparison with simulations, we can turn to $r_s$. We are particularly interested in comparing our recovered value of $r_s\sim 0.85$ --- which is indicative of a strong correlation between $\mathrm{t}_{\mathrm{halo}}$ and M$_*$ at fixed M$_\mathrm{H}$ --- with expectations from semi-empirical models. For this purpose, we analyzed data for the SHMR from the UniverseMachine. We calculated $\mathrm{t}_{\mathrm{halo}}$ following the same definition outlined in Section \ref{3.1} and computed $r_s$ between $\mathrm{t}_{\mathrm{halo}}$ and M$_*$ at fixed M$_\mathrm{H}$. We found $r_s$ to peak at M$_\mathrm{H}\sim10^{11}-10^{12}~h^{-1}$~M$_{\odot}$ with values of $r_s \sim 0.5-0.6$. We conclude that semi-empirical models like the UniverseMachine predict a correlation between $\mathrm{t}_{\mathrm{halo}}$ and M$_*$ that is only slightly weaker than what we inferred from SDSS.

So far, we have discussed how variations in halo formation time can affect the M$_*$ (this work) and stellar ages (\citealt{oyarzun2022}) of central galaxies (Figure \ref{fig5}). We should highlight that the stellar population variations that we identified in \citet{oyarzun2022} also extend to chemical content of galaxies. Besides being older, centrals with high M$_*$-to-M$_\mathrm{H}$ ratios also feature low [Fe/H] and high [Mg/Fe]. The natural interpretation for these trends is that centrals with high-M$_*$ for their M$_\mathrm{H}$ not only form their stars early, but that they also assemble them in rapid timescales, resulting in high [$\alpha$/Fe] (e.g. \citealt{faber-jackson1976,faber1985,thomas2005}). Variations in the star-formation timescale may also be connected with variations in secondary halo properties, in particular with halo formation time and halo concentration. At fixed M$_\mathrm{H}$, older and higher-concentration halos are thought to bind the stellar and gaseous contents of centrals more strongly (e.g. \citealt{wechsler2002,hearin2016}). These results --- along with the work by \citet{winkel2021,donnan2022,zhou2024} --- provide tentative observational evidence for a correlation between secondary halo properties and the quenching epochs and the metal content in galaxies (as seen in simulations; e.g. \citealt{booth-schaye2010,matthee2017,montero-dorta2020,montero-dorta2021,hasan2023a}). 

Our studies of how $\delta_{10}$ and the stellar populations of galaxies vary within the SHMR were in great part enabled by publicly available datasets of high statistical power (i.e., SDSS and MaNGA). SDSS is the largest publicly available spectroscopic survey of galaxies in the nearby Universe, providing the necessary statistics for dissecting the SHMR, the required spectra for accurately measuring stellar masses, and the deep ancillary imaging needed for precise estimation of the halo mass. At the same time, MaNGA is the largest spatially resolved spectroscopic survey of nearby galaxies. In terms of spectroscopic constraining power, MaNGA surpasses even SDSS (see \citealt{oyarzun2022}), allowing for the most precise measurements of stellar age and chemical abundances possible in the SHMR of nearby galaxies. This will soon change, with surveys like the Bright Galaxy Sample of the DESI Experiment (\citealt{DESI}) spearheading the next generation of datasets that will enable detailed characterization of the galaxy-halo connection at low redshifts.

\section{Summary}
\label{6}

We used the group catalog from \citet{tinker2021,tinker2022} to study the stellar-to-halo mass relation for red central galaxies from SDSS across the halo mass range M$_\mathrm{H} = 10^{11}-10^{15}~h^{-1}$~M$_{\odot}$. We showed that the stellar masses of red central galaxies vary not only with the mass of the dark-matter halo, but also with the number density of galaxies in large scales ($R=10~h^{-1}$~Mpc). Our preferred interpretation for this result is that the stellar masses of central galaxies depend on both halo mass and halo formation time. This interpretation is motivated by our current cosmological picture, in which high density regions of the large scale structure assemble earlier in cosmic history.

To quantify the relation between the large-scale number density of galaxies and their halo formation time, we exploited the MDPL2 dark-matter-only cosmological simulation (\citealt{prada2012}). With the halo formation times in hand, we constructed a model that reproduces differences in the observed number density of galaxies via two parameters: (1) random errors in the halo masses estimated by the group catalog and (2) the correlation between halo formation time and stellar mass at fixed halo mass (i.e., the galaxy assembly bias signal). We quantified the model posterior with SDSS and found that errors in the halo masses from the group catalog are $\sim 0.15$~dex (in agreement with the expected $\lesssim 0.2$~dex errors from the group finder algorithm; \citealt{tinker2022}) and that galaxy assembly bias is necessary for explaining our results at $>5\sigma$ significance.

Our best-fit model indicates that variations in the stellar masses of red central galaxies at fixed halo mass correlate with differences in halo formation time of $\sim 1-3$~Gyr. Differences of this magnitude are remarkably consistent with differences in the star-formation histories of galaxies at fixed halo mass, which we measured previously in \citet{oyarzun2022}. Taken together, these results are indicative that old halos (at fixed halo mass) are associated with more massive, older, and more rapidly assembled central galaxies.

We also considered how systematic errors in the group finder algorithm could fabricate the signal that we observe in the galaxy number densities. While systematic and random errors are likely present, they alone cannot reproduce the signal. This is because we still detect a similarly significant difference in the galaxy number densities after substituting M$_\mathrm{H}$ for total group luminosity and M$_*$ for galaxy luminosity.

\section{Data Availability}

Underlying data in this article will be shared upon reasonable request to the corresponding author.

\medskip

We thank the reviewer of this paper for their insightful comments and suggestions. G.O. acknowledges support by the National Science Foundation through grants AST-2107989 and AST-2107990. This work made use of Astropy\footnote{http://www.astropy.org}, a community-developed core Python package and an ecosystem of tools and resources for astronomy \citep{astropy:2013, astropy:2018, astropy:2022}. Funding for the Sloan Digital Sky Survey (SDSS) has been provided by the Alfred P. Sloan Foundation, the Participating Institutions, the National Aeronautics and Space Administration, the National Science Foundation, the U.S. Department of Energy, the Japanese Monbukagakusho, and the Max Planck Society. The SDSS Web site is http://www.sdss.org/. The SDSS is managed by the Astrophysical Research Consortium (ARC) for the Participating Institutions. The Participating Institutions are The University of Chicago, Fermilab, the Institute for Advanced Study, the Japan Participation Group, The Johns Hopkins University, Los Alamos National Laboratory, the Max-Planck-Institute for Astronomy (MPIA), the Max-Planck-Institute for Astrophysics (MPA), New Mexico State University, University of Pittsburgh, Princeton University, the United States Naval Observatory, and the University of Washington. The MultiDark Database used in this paper and the web application providing online access to it were constructed as part of the activities of the German Astrophysical Virtual Observatory as result of a collaboration between the Leibniz-Institute for Astrophysics Potsdam (AIP) and the Spanish MultiDark Consolider Project CSD2009-00064. The Bolshoi and MultiDark simulations were run on the NASA’s Pleiades supercomputer at the NASA Ames Research Center. The MultiDark-Planck (MDPL) and the BigMD simulation suite have been performed in the Supermuc supercomputer at LRZ using time granted by PRACE.

\bibliography{sample631}{}
\bibliographystyle{aasjournal}



\end{document}